\documentstyle[12pt,moriond,psfig]{article}

\begin{document}

\heading{A Numerical Study of galaxy properties from cosmological
  simulations with star formation}

\author {G. Yepes, D. Elizondo and Y. Ascasibar } {Grupo de  Astrof\'{\i}sica,}{ Universidad Aut\'onoma de Madrid, Spain }

\begin{moriondabstract}
From a  set of 3d cosmological hydrodynamical simulations which
incorporate a self-consistent model  of  the star-gas interactions,
 we have been able to obtain a statistically significant
 sample of galaxy-type  halos with observational properties,  like colors
and luminosities.  Using this data-base, we
have studied  general relations of galaxies, such as  
the Tully-Fisher relation, luminosity functions, environmental
dependences, or  cosmic   star-formation density.  The main motivation
of our analysis was to investigate the  influence
 of   cosmological  conditions (i.e. large-scales) and
 (short-scale)  baryonic processes on the   observational  properties
of the resulting {\em numerical} galaxies.
\end{moriondabstract}

\section{Introduction}
Our  numerical code is a combination of a Particle-Mesh (PM)
 Poisson solver and an  Eulerian PPM hydrodynamical code.  
A detailed description of the N-body/hydro code and 
the equations governing the  star- gas interactions  can be found in Ref.  \cite{yk3} (YK$^3$). Here we briefly summarize their  main features. 
In our numerical model,
 matter is treated as a multi-phase fluid with different
physical processes acting between them. The gas is
treated as  two separate phases, depending on its temperature: 
{\sl Hot Gas }($T_h > 10^4 $K) and {\sl  Cold Gas Clouds} ($T_h < 10^4
$K) from which stars are formed.
In each
resolution element,  the amount of cold gas,  $m_{\rm cold}$, 
 capable of producing stars is regulated by the mass of the hot gas,
 $m_{\rm hot}$,  (that can cool on the time scale $t_{\rm cool}$)
, by the rate of forming
new stable stars, and by supernovae (SN), which heat and evaporate
cold gas. The supernovae formation rate is assumed to be proportional
to the mass of cold gas: $\dot m_{\rm SN}=\beta m_{\rm
cold}/t_*$, where $t_*=10^8$yr is the time scale for
star formation, and $\beta$ is the fraction of mass 
that explode as supernovae ($\beta=0.12$ for the Salpeter
IMF). Each $1M_\odot$ of supernovae dumps $4.5\times 10^{49}$ ergs of
heat into the interstellar medium {\it and} evaporates a mass $A\cdot M_\odot$
of cold gas.  Small values of
$A$ imply  large reheating of hot gas and small evaporation, which
makes the gas to expand due to the large pressure gradients.
The feedback parameter $A$ is taken to be large ($A\sim 200$) 
resulting in low efficiency of converting cold gas into
stars. 
Chemical enrichment due to supernovae  is also taken into
account in the following way:
 We assume
solar composition for the gas in regions where star formation has taken
place. In regions where no stars are present, we assume that  the gas
has primordial composition. The gas then cools  with cooling rates,
$\Lambda(T_h)$ corresponding to either primordial or solar plasmas. 
 In order to mimic the effects
of photoionization by quasars and AGNs, the gas with overdensity less
than 2 was kept at a constant temperature of 
$10^{3.8}$K (see e.g. Refs \cite{gs}, \cite{pmk}).

\section{Description of simulations}
The purpose of the simulations was to obtain a sufficiently large
catalog of ``numerical galaxies'', permitting reliable, statistically
significant comparisons with observational quantities.
To this end, a
set of 11 simulations were performed for each of the CDM, $\Lambda$CDM
($\Omega_\Lambda = 0.65$), and BSI models \cite{kat3}.
COBE
normalization was taken; baryon fractions were compatible with nucleosynthesis
constraints \cite{nucleo2}  ($\Omega_B=0.051$ for BSI and CDM,
$\Omega_B=0.026$ for $\Lambda$CDM ).
The box size was chosen to be  5.0 Mpc,
 but with different Hubble constants given by $h=0.7$ for the
$\Lambda$CDM model and $h=0.5$ for the CDM and BSI simulations. 
  The simulations  were performed at the
CEPBA {\em (Centro Europeo de Paralelismo de Barcelona\/}) with 128$^3$
particles and cells (i.e., 39 kpc cell width).
Effects of resolution  have been checked by re-running 2 of the simulations
with $256^3$ cells and particles. It 
 did not result in significant changes in global parameters (mass,
 luminosity)  of  galaxies.
To test the effects of supernovae feedback on the  final observational
properties of galactic halos, we have rerun 6 of the 11 simulations for
each cosmological model, with different feedback parameter: $A=50$,
(strong gas  reheating) and $A=\beta=0$, (no  reheating or mass
transfer). 
A detailed analysis of these simulations  can be found elsewhere (\cite{nosotros}-\cite{nosotros2})

 \begin{figure}[bth]
\centerline{\psfig{file=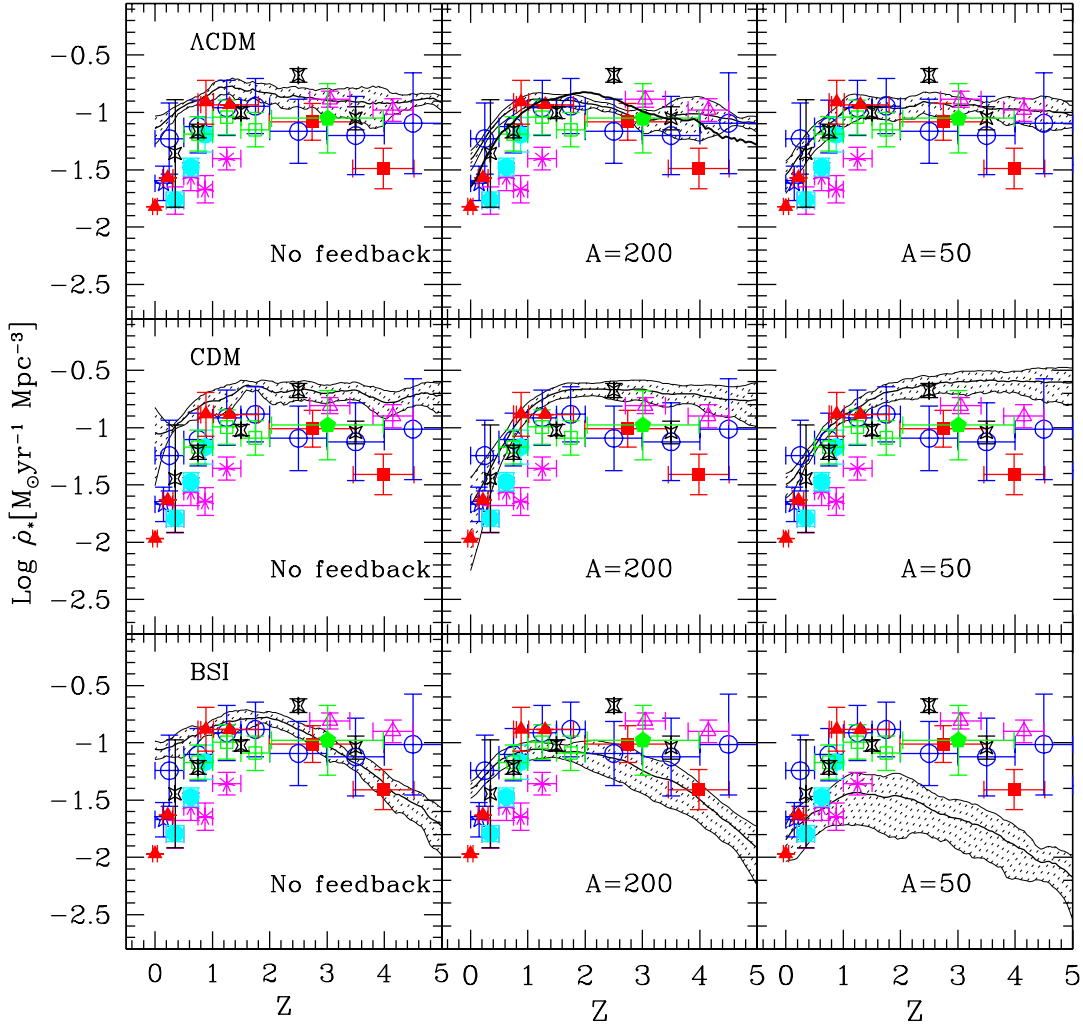,height=15cm}}
\vspace*{-0.8cm}
\caption{Comoving star formation density as a function of
  redshift. Solid line represent the mean over  6 different
  realizations. Dashed areas are the  1 $\sigma$  deviation from 
the mean. The thick solid line in the
$A=200$, $\Lambda$CDM panel show results for  the  12 Mpc box simulation
with same resolution than the 5 Mpc ones.  Observational points  are
collected from literature and they have been rescaled  according to 
to the  cosmology  corresponding to each model. 
    \label{fig:fig1}}
\end{figure}

More recently, we have completed a $\Lambda$CDM  simulation with $300^3$ 
particles in a 12 Mpc box  (i.e. same resolution than in a 5 Mpc
box). This simulation was done to check for possible effects due to the
lack of long wavelenghts in the initial power spectrum.

\section{Results}
From the data base of galaxy-type halos extracted from the
abovementioned simulations we have  studied the Tully-Fisher (TF) relation
in different bands (B, R, I) as well as the luminosity function in B
and K \cite{nosotros}.

The luminosity functions in the $B$ and $K$ bands are quite
sensitive to supernova feedback.
We find that the slope of the faint end ($-18 \leq M_B \leq -15$) of the $B$-band luminosity function 
 is $\alpha  \approx$  $-1.5$ to  $-1.9$. This slope is steeper than
the Stromlo-APM estimate, but in rough agreement 
with the recent ESO Slice Project  \cite{zuca}.
The galaxy catalogs derived from our hydrodynamic simulations lead
to an acceptably small scatter in the theoretical TF relation
amounting to $\Delta M =0.2-0.4$ in the $I$ band, and increasing
by 0.1 magnitude from the $I$-band to the $B$-band.  Our results
give strong evidence that the tightness of the TF relation cannot
be attributed to supernova feedback alone.  
However, although eliminating  supernova feedback affects the scatter only
moderately
($\Delta M = 0.3-0.6$), it 
does influence the slope of the TF relation quite sensitively. 
With supernova feedback, $L\propto V_c^{3-3.5}$ 
(the exponent depending on the degree of feedback). Without
it, $L\propto V_c^{2}$ as predicted by the virial
theorem with constant $M/L$ and radius independent of luminosity.

In Figure \ref{fig:fig1} we show the redshift evolution of the
comoving star-formation density , $\dot{\rho_\star}$, of our
simulations,  together with a compilation  of the most recent 
 observational estimates (see \cite{co},\cite{madau99}, \cite{pasc},
 \cite{sawi}, \cite{st}, \cite{yan}) derived from the UV luminosity
 density,  corrected from dust extinction, following   Madau's prescription \cite{madau}.

In the low evolved BSI simulations,
 the effects of SN  feedback  on $\dot{\rho_\star}$  are  striking 
at all redshifts,
 while in the CDM and $\Lambda$CDM simulations,
feedback effects   become  significant   only at  $z \lsim
1$: Simulations with  SN feedback  show a much sharper decline of
$\dot{\rho_\star}(z)$, which  is a reflection of the decline of
the SFR inside the bright galactic halos at low $z$ as a consequence of
the higher temperature of the hot gas \cite{nosotros2}.

 In higher normalized CDM and $\Lambda$CDM simulations,    $\dot{\rho_\star}(z)$ is 
almost flat  or  slightly declining     at   $z \sim 1-5$. This is  in good 
 agreement with  the recent  observational  data
  when  correction from dust extinction is considered.  In the
 $\Lambda$CDM--A=200 panel, we also show  the  $\dot{\rho_\star}(z)$
 computed   in the 12 Mpc simulation box.  As can be seen, 
it is consistent  with  the  results from the  5 Mpc simulations.
Morover,  they are also in fairly good agreement with 
estimates  of   $\dot{\rho_\star}(z)$  computed  from  
large-scale (100 h$^{-1}$Mpc)  $\Lambda$CDM 
hydrodynamical simulations \cite{cen},  with an
analytical prescription for the SFR  inside  galactic halos.

\section{Conclusions and future work}
Despite the success of our  model to reproduce some  of the
properties of real galaxies, it is nevertheless a   simplified model of the
complex star-gas interactions. UV photoionization and
chemical enrichment are treated in a phenomenological way.  These
effects could be  very important, and could change the evolution
of the gas component. 
 The advantage of using eulerian PPM hydrodynamics is that it
 is simpler to advect  metals as a new ``phase'' of the gas density,
 which would change the local cooling rates of the gas.
 Stars  will then form   with different metallicities 
 and their luminosities can be computed by  the new generation of 
population synthesis models. 
 UV radiation from the stars is  another important effect that has not
 yet been fully explored.  Our purpose 
 is to make a self-consistent modeling  of
 the photoionization of the gas from UV flux coming from the stars
 generated  in the simulations.  This will constitute another 
feedback  mechanism in  our star-gas model.

 The main  goal one wants  to achieve  in  a cosmological simulation is
 to resolve the internal structure of individual galaxies  formed in
 volumes that are large enough  to allow a reliable realization of the initial
 power spectrum.  New numerical algorithms  based on 
{\em Adaptive Mesh Refinement (AMR)} techniques   are starting to be
considered  one of the most promising ways  to  pursue  this goal. 
 But higher resolution does not necessarily mean better
results, unless the most important  physical processes acting 
at the scales of interest have been  included in the simulation. Our
work follows this premise.
 On one hand it is necessary to explore
many approximations  to properly model the complex physics of star
formation and star-gas interactions \cite{asca}.
 On the other hand, new AMR methods
for  gravity \cite{art}  and   hydrodynamics \cite{ppma} 
have already been  developed.  The next logical   step
 is to put together the different pieces
  and build  a new  generation of  numerical simulations
to study  galaxy formation from cosmological initial
conditions.  In this regard, the  data base  obtained from
simulations  performed with our present numerical code   (YK$^3$)
  will be  very useful as a testbed  for the new   models  we are
currently   developing.



%
\vfill
\end{document}